%% file: 0_Main.tex
\documentclass[a4paper,11pt]{article}
\pdfoutput=1 % if your are submitting a pdflatex (i.e. if you have
\usepackage{jinstpub} % for details on the use of the package, please
\RequirePackage{lineno}\newdimen\linenumbersep\linenumbersep=2pt

\title{Correlated Single- and Few-Electron Backgrounds Milliseconds after Interactions in Dual-Phase Liquid Xenon Time Projection Chambers}

\author{Abigail Kopec,}
\author{Amanda L. Baxter,}
\author{Michael Clark,}
\author{Rafael F. Lang,}
\author{Shengchao Li,}
\author{Juehang Qin,}
\author{Riya Singh}

\affiliation{Department of Physics and Astronomy, Purdue University, West Lafayette, Indiana 47907, USA}
\emailAdd{kopec@purdue.edu}

\abstract{We characterize single- and few-electron backgrounds that are observed in dual-phase liquid xenon time projection chambers at timescales greatly exceeding a maximum drift time after an interaction. These instrumental backgrounds limit a detector's sensitivity to dark matter and cosmogenic neutrinos. Using the $\sim 150$~g liquid xenon detector at Purdue University, we investigate how these backgrounds, produced after 122~keV $^{57}$Co Compton interactions, behave under different detector conditions. We find that the rates of single- and few-electron signals follow power-laws with time after the interaction. We observe linearly increasing rates with increased extraction field. The relationship of the rates in the single-electron background with increased drift field is unclear. Normalizing the rates to the primary interaction's measured ionization signal, the rates increase linearly with the depth of the interaction. We test the hypothesis that infrared photons (1550~nm) would stimulate and accelerate electron emission via photodetachment from impurities, but find that even 1 Watt of infrared light fails to reduce these backgrounds. We thus provide a characterization that can inform background models for low-energy rare event searches.
}


\begin{document}
\maketitle
\flushbottom

\input{1_Intro.tex}
\input{2_TheTPC.tex}

\input{3_Electron_Trains.tex}

\input{4_Conclusion.tex}
\input{Acknowledgements.tex}
\input{asterix.tex}

\end{document}

%% file: 1_Intro.tex
\section{Introduction}
\label{intro}

Liquid xenon Time Projection Chambers (TPCs) are excellent particle detectors for rare event searches due to their signature low backgrounds. They are particularly well-suited to searching for Weakly Interacting Massive Particles (WIMPs) at masses above a few GeV/c$^2$~\cite{Aprile:2018dbl,Akerib:2016vxi,Wang:2020coa}. Beyond WIMPs, these versatile detectors can be used to look for neutrinoless double-beta decay~\cite{Agostini:2020adk,Anton:2019wmi}, supernova and solar neutrinos~\cite{Newstead:2020fie,Aalbers:2020gsn,Lang:2016zhv}, and other dark matter candidates. Through some channels, they are sensitive to dark matter masses well below 1~GeV/c$^2$~\cite{Essig:2011nj}, which is the goal of the dedicated LBECA TPC~\cite{Bernstein:2020cpc}. Signatures of these dark matter candidates could appear, for example, as Electronic Recoils~\cite{Aprile:2020tmw}, via the Migdal Effect~\cite{Aprile:2019jmx,Akerib:2018hck}, or in isolated ionization signals~\cite{Aprile:2019xxb,Aprile:2016wwo}. With such signals expected to manifest near the detection threshold, understanding instrumental backgrounds to low-energy interactions is of paramount importance.  

The lowest-energy interaction detectable in a liquid xenon TPC would cause a single-electron ionization signal, requiring $\sim$15~eV~\cite{Boulton:2017hub,Akerib:2017hph} for electronic recoils and $\sim$250~eV~\cite{Lenardo:2019vkn} for nuclear recoils. The light collection efficiency is typically low for detecting individual scintillation photons, but electrons give large signals and are detectable with high efficiency~\cite{Burenkov:2009zz,Santos:2011ju,Angle:2011th,Aprile:2013blg,Akerib:2017vbi,Edwards:2017emx,Xu:2019dqb}. As of now, background rates for small ionization signals up to five electrons far exceed current background models and are not well understood~\cite{Aprile:2019xxb}. This paper seeks to characterize such backgrounds in order to improve these detectors' sensitivities to lower mass dark matter candidates and cosmogenic neutrinos.

Single electrons correlated in time with a previous high-energy interaction have been observed out to times much longer than the maximum drift time of a free electron in the detector, and are sometimes known as ``electron trains"~\cite{Angle:2011th,Akimov:2016rbs,Akerib:2020jud}. The single-electron signal rates decrease with time after the primary interaction, and reach a lower average rate when there are fewer high-energy interactions~\cite{Santos:2011ju,Akimov:2016rbs}. Sorensen and Kamdin identified two exponential components to the rate evolution~\cite{Sorensen:2017kpl}, but a power law appears to be a better fit~\cite{Akimov:2016rbs}. Data reported by LUX is consistent with the power law, although no fit was performed~\cite{Akerib:2020jud}.

\textit{A priori} models for electron emission from the liquid surface have failed to satisfactorily describe data~\cite{Gushchin:1979,Gushchin:1982}. Empirically, a few qualitative hypotheses for spurious single-electron signals have been offered~\cite{Burenkov:2009zz,Santos:2011ju,Aprile:2013blg,Akimov:2016rbs,Bodnia:2021flk}. Two main conjectures to explain position- and time- correlated delayed electron emission of the electron trains have emerged: electrons from the primary interaction are delayed either at the liquid-gas interface~\cite{Sorensen:2017ymt,Akimov:2012zz}, or on electronegative impurities~\cite{Sorensen:2017kpl,Bodnia:2021flk}. 

Regarding the liquid-gas interface, accounting for electrons taking multiple attempts to tunnel out of the liquid offers a potential solution~\cite{Sorensen:2017ymt}. When a slight tilt is applied to the detector such that the liquid surface is no longer perpendicular to the extraction electric field, rates of spurious electrons decrease~\cite{Akimov:2012zz} and the emission locations appear to drift to where the liquid electric field is greatest~\cite{Akimov:2016rbs}.

On the other hand, the absolute number of these delayed signals decreases when the xenon purity improves~\cite{Aprile:2013blg}, and increases with interaction depth~\cite{Akerib:2020jud}. These observations favor the impurity hypothesis, since purity seems to have an effect, and interactions deeper in the detector lose more electrons to impurities in the longer drift column~\cite{Akerib:2020jud}. There are also reports of decreased rates of single electrons with increased drift field~\cite{Bodnia:2021flk}. It is proposed that different electric fields in the liquid affect the physics of trapping and releasing electrons on impurities~\cite{Sorensen:2017kpl}. Beyond xenon TPCs, DarkSide-50 has observed that the number of small S2 signals in their dual-phase argon TPC decreased with better purity~\cite{Agnes:2018ves}. However, these studies have focused primarily on single-electron signals, and the rates do not have a simple dependence on purity~\cite{Akimov:2019ogx}.

In this paper, we investigate single- to five-electron signals following 122 keV $^{57}$Co Compton scatters as primary interactions, and how their rates up to a few milliseconds depend on detector configurations. In Section \ref{TheTPC}, we describe our Purdue University research TPC. In Section \ref{results}, we present our analysis of delayed ionization signals, which we discuss in Section \ref{conclude}.

%% file: 2_TheTPC.tex
\section{The ASTERiX Detector}
\label{TheTPC}

In a dual-phase liquid xenon TPC, a particle scattering with either a xenon nucleus or an electron transfers momentum to that target particle, which loses its kinetic energy by exciting xenon atoms in its recoil path. Therefore, the kinematics of the interaction can be reconstructed based on the total number of excited atoms. Many of the atoms promptly de-excite, producing prompt scintillation photons seen by the photosensors (S1). The rest of the atoms lose their excited electrons, which drift to the top of the detector in the electric field and are extracted into the gas. There, in the amplification region, each electron interacts with the gaseous xenon, producing scintillation light in proportion to the number of extracted electrons (S2). The photon hit-pattern in the top photosensor array for the S2 gives horizontal positioning coordinates (x,y) and the drift time of the electrons between the S1 and the S2 gives the depth in the detector (z) of the interaction.

\begin{figure}[htb]
\centering
\includegraphics[scale=0.35]{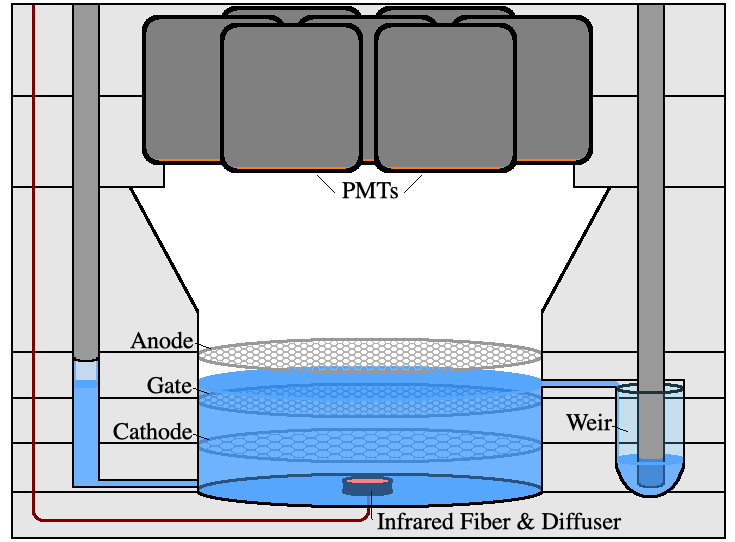}
\caption{A cross-sectional view of ASTERiX. Three etched electrode meshes, separated by 1~cm each, provide the electric fields for the detector, which is 7.5~cm in diameter. Seven PMTs in a top array collect scintillation light. The liquid xenon level is maintained 2.5~mm above the gate with a weir to recirculation. A fiber-coupled diffuser shines infrared light into the chamber from the bottom.}
\label{fig:ASTERiX}
\end{figure}

\textbf{A} \textbf{S}mall \textbf{T}PC for \textbf{E}xperimental \textbf{R}esearch \textbf{i}n \textbf{X}enon (ASTERiX) is the small-scale detector at Purdue University, as shown in Figure \ref{fig:ASTERiX}. ASTERiX has seven 1-inch square Hamamatsu R8520 photomultiplier tubes (PMTs) in a top array to collect xenon scintillation light from a cylindrical active volume. The PMTs are negatively biased to -680~V, but positioned 5~cm above the positively biased anode, which avoids electric discharge between the anode and PMTs. The entire detector is made from highly reflective PTFE. The target volume is 7.5~cm in diameter with a 1~cm high drift region between the cathode and the gate electrodes, containing $\sim$150~g of liquid xenon. A weir maintains the liquid xenon level 2.5~mm above the gate, leaving a 7.5~mm gas amplification region below the anode. The three electrodes are stainless steel meshes, etched with a hexagonal pattern. Liquid xenon spilling into the weir is pumped out during constant recirculation, evaporating to gas and passing through a hot zirconium getter before re-entering the detector near the coldhead. Warm xenon condenses on the coldhead, drips into a funnel, and feeds back into the detector volume. In the current configuration, infrared light can illuminate the chamber through a fiber-coupled optical diffuser below the cathode. 

The TPC is positioned at the bottom of a meter-high stainless steel vacuum-insulated cryostat. The meter distance thermally isolates it from a single-walled region with the electric feedthroughs. A 4-inch thick wall of lead bricks surrounds the cryostat at the detector location to shield it from cosmic rays in the above-ground location, which drops the event trigger rate from $\sim$200 to $\sim$10~Hz. A 1000~Bq $^{57}$Co source was placed inside the lead fort but at a distance from the outside of the cryostat such that the trigger rate was $\sim$20~Hz.

Data from the PMTs is collected with a CAEN V1724 digitizer. For this study, a 2-PMT trigger coincidence was required and the trigger threshold was set high to preferentially trigger on the 122~keV $^{57}$Co Compton scatter interactions. The trigger window was set to the digitizer maximum of 524,288 ten-nanosecond samples to achieve just over 5~ms after each primary interaction. We used the zero-length encoding feature to reduce data and baseline noise well below a single photon signal. In this configuration, the digitizer's maximum trigger rate was roughly 11~Hz due to significant dead time for processing. Despite this limitation, the digitizer triggered on the desired $^{57}$Co interactions and collected data for milliseconds immediately afterward.  

This data was then fed into an adapted version of the Processor for Analyzing XENON (PAX)~\cite{PAX_github}. This software can identify individual S1 and S2 signals and calculate relevant quantities, such as how many photoelectrons (PE) the PMTs observed for each signal, the ($x,y$)-position of S2s, when the signals happened, and the duration or width of the signals in time. It also finds interactions by matching likely S1s and S2s and finding the drift time and $z$-position. With this information, a thorough analysis of few-electron s2s after large energy depositions is performed.

%% file: 3_Electron_Trains.tex
\section{Delayed Ionization Signals}
\label{results}

\begin{figure}[htb]
\centering
\includegraphics[scale=0.345]{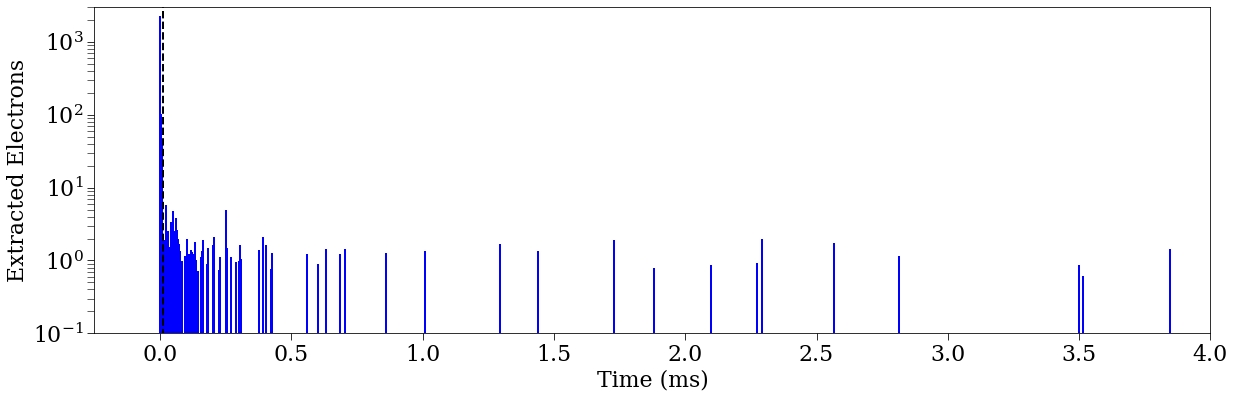}
\caption{An example $^{57}$Co event, with electron signals appearing for milliseconds afterward. This graph shows the central time locations of all reconstructed S2s in the event window and their electron multiplicity. The primary interaction's S2 is at 0.0~ms. The maximum drift time of electrons in ASTERiX is 10~$\mu$s, represented by the dashed line.}
\label{fig:event}
\end{figure}

The default detector settings were chosen to have +5~kV applied to the anode, -5~kV applied to the gate and -5.5~kV applied to the cathode. This gives a drift field of 500~V/cm and an extraction field of 5.9~kV/cm in the liquid and an 11~kV/cm amplification field in the gas. With such an amplification field, the single electron gain is $15.3 \pm 0.1$~PE. The maximum drift time is $\sim$10~$\mu$s. With the data acquisition settings described in Section~\ref{TheTPC}, we took data for 15-minutes, which resulted in roughly 5,000 event windows of over 5~ms each. Our data clearly show the previously described electron trains that continue for long times after a large energy deposition. An example event is shown in Figure~\ref{fig:event}.  

Since we are concerned with S2s up to five electrons, all S2s greater than six electrons, 90 PE, were considered primary interactions. Consecutive large S2s within a maximum drift time were grouped together into a single primary interaction S2, and their median timestamp, weighted by size, was taken as the combined S2's time. This correction was necessary for S2s incorrectly split by the processor, and multiple-scatter events. It does not affect the relative behavior of backgrounds several tens of microseconds later. Smaller S2s were matched in time with the most recent primary S2. With all S2s matched, we implemented an exponential overlap cut in time, based on the primary S2s' sizes, to select primary interactions with the least contamination from previous primary interactions' electron trains. A primary S2 event is accepted if its size is larger than 10\% of the previous primary S2's size reduced exponentially according to a time constant of 1~ms. We finally selected from these the true $^{57}$Co interactions and their electron trains, based on the S1 and S2 energies and their corresponding drift times. A typical $^{57}$Co event would produce $\sim 10^4$ electrons, but a short electron lifetime due to poor purity exponentially decreases the measured S2 size based on the interaction's depth.
 
We expect photoionization electron backgrounds from the TPC walls, impurities, and electrodes after the bright primary S2 for at least a drift time~\cite{Aprile:2013blg}, potentially continuing afterward with more photoionization induced by the first photoionization S2s, but dropping exponentially. To avoid this known background, we focused on the S2s that came at least three times the maximum drift time after the primary interaction's S2. For the default detector settings, this was set at 30~$\mu$s. 

\begin{figure}[htb]
\centering
\includegraphics[scale=0.5]{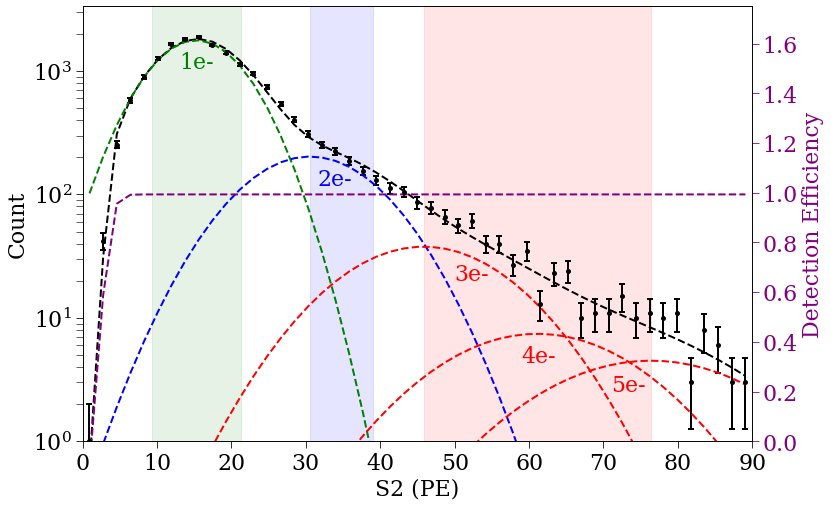}
\caption{The spectrum of small S2s in the electron trains following over 30~$\mu$s after accepted $^{57}$Co primary interactions. Gaussians were fit to each electron multiplicity, and the shaded regions denote the S2 energy ranges used for the singe-electron (green), double-electron (blue), and 3-5 electron (red) populations.}
\label{fig:spectrum}
\end{figure}

Figure~\ref{fig:spectrum} shows the spectrum of low-energy S2s and the fit of the single- and few-electron signals. Then, to understand the rate behavior of purely single-electron, double-electron, or 3-5 electron S2 populations at these long times after a primary interaction, we selected the S2s in the shaded regions. The final rates account for the purity of the populations and efficiencies of the population cuts.

\begin{figure}[htb]
\centering
\includegraphics[scale=0.34]{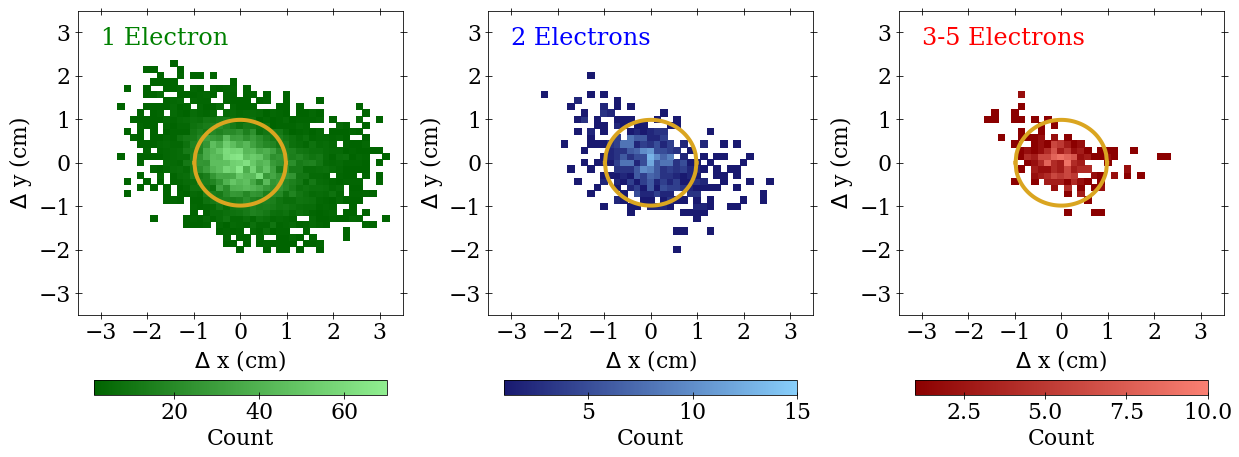}
\caption{The displacement of signals from the primary interaction location for single-electron, double-electron, and 3-5 electron populations (left to right). The gold circle denotes the radius 9.9~mm, containing 80\% of the position-correlated single-electron populations, and $\sim$70\% of all electron train S2 signals.}
\label{fig:xyloc}
\end{figure}

The positions of the selected S2s relative to the location of their corresponding primary interactions are shown in Figure~\ref{fig:xyloc}. Based on the detector geometry, the $^{57}$Co source location, and position reconstruction tendencies, we simulated pairs consisting of a random possible position and a random primary position to determine the expected displacement distribution for position-uncorrelated backgrounds. This simulated data was used to construct a coordinate transformation for the displacement values in data such that the uncorrelated background distribution would be flat and easily subtracted from the position-correlated signals. We applied the coordinate transformation to the single electrons, which had the most statistics. With the flat background identified and subtracted, the radius containing 80\% of position-correlated single electrons is 9.9~mm. We applied this radius to the double and 3-5 electron populations. About 70\% of all populations of small S2s are found to be position-correlated in this radius.
 
\begin{figure}[htb]
\centering
\includegraphics[scale=0.45]{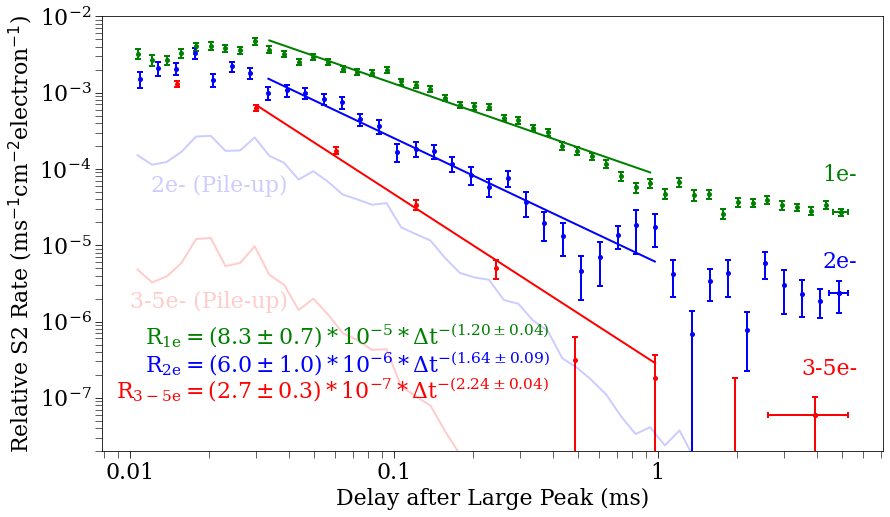}
\caption{The evolution of the rates of position-correlated single (green), double (blue) and 3-5 (red) electron signals after a $^{57}$Co interaction with time. They are normalized by the position-correlated area of 9.9~mm radius, and the number of electrons produced in the primary interaction (corrected for electron lifetime and drift time). The horizontal error-bars of the last data point in each series illustrate the bin ranges. The expected contribution from pile-up is indicated for the double (light blue) and 3-5 (light red) electron signals.}
\label{fig:rateplot}
\end{figure}

 To recapitulate: we avoided overlap from previous electron trains through the exponential overlap cut, and selected the electron trains of $^{57}$Co primary interactions based on S1, S2, and drift time variables. Within the electron trains, we selected the purest populations for each electron multiplicity within specific energy bounds, and the position-correlated small S2s within 9.9~mm displacement from the primary interaction location. The rates of these signals are finally calculated, accounting for varying livetime windows between primary interactions, or to the ends of digitized event windows. The rates are also normalized by the number of electrons produced in the primary S2, and are adjusted per detector area within the 9.9~mm radius, due to incomplete circular areas at the edges of the detector. They are lastly scaled by the purity and efficiency of the population selections previously mentioned. The rates for different electron multiplicities are shown in Figure~\ref{fig:rateplot}. We typically did not have the statistics to investigate the position-uncorrelated backgrounds independent of the position-correlated effects.

A power law fit $R = A \cdot t^{b}$ using a maximum likelihood~\cite{Goldstein_2004} was applied between 30~$\mu$s and 1 ms for each electron multiplicity population, with the fit results listed in the figure. Despite describing the data much better than an exponential (or even the sum of up to three exponentials), a power law may not be the best model, indicated by the apparent curvature in the single-electron rates. Regardless, the single-electron rate exponent $b$ of -1.20 $\pm$ 0.04(stat) fits with the roughly -1 exponents previously observed in other detectors~\cite{Akimov:2016rbs}. 

The amplitudes and powers of the double and 3-5 electron populations indicate that the mechanism producing single-electrons is also able to produce signals with higher electron multiplicities that are not due to pile-up or contamination from single-electron signals. Regarding pile-up, the coincidence window required for two independent single-electrons to be reconstructed into a double-electron S2 is less than a microsecond. For a Poisson process with a given rate $R$ of single-electron signals, the probability of observing a coincidence time window $t$ with n-electron pile-up S2s is given below in Equation~\ref{eq:pois}.
\begin{equation}
    P(n|R \cdot t)=\frac{(R\cdot t)^n}{n!}e^{-R\cdot t}
    \label{eq:pois}
\end{equation}
Therefore, the rate of pile-up $R_n$ to form an n-electron S2 is related to the rate of single-electron S2s $R_1$:
\begin{equation}
    R_n = \frac{(R\cdot t)^n}{n!}R\cdot e^{-R\cdot t} = \frac{(R\cdot t)^{n-1}}{n!}R_1
    \label{eq:rat}
\end{equation}
When the total rate $R$ in a sufficiently small coincidence window has an expectation value less than one, the single-electron rate dominates $R$. In the case of the double-electron S2s from pile-up, their rate $R_2$ can be approximated in terms of the single-electron rate:
\begin{equation}
    R_2 = \frac{R_1^2 \cdot t}{2}
    \label{eq:doub}
\end{equation}
By this expectation for pile-up, and taking into account the normalization factors for position-correlation area selection within a 9.9~mm radius and the number of electrons produced in the primary interaction of typically $10^4$ electrons, the double-electron S2 contribution from single-electron pile-up is an order of magnitude below the measured double-electron signal rates as shown in Figure~\ref{fig:rateplot}. Additionally, the double-electron power $b$ would necessarily be twice that of the single-electrons, since the pile-up rate roughly scales as $R_1^2$, which is not consistent with our observations.

Regarding contamination, the single-electron Gaussian tail contamination in the double-electron population is less than 10\%. Since the double-electron rates are within an order of magnitude of the single-electron rates and the powers are different, we conclude that these are true double-electron signals, and not from single-electron contamination or pile-up. The same arguments apply to the 3-5 electron populations, indicating some mechanism that can produce delayed, position-correlated signals with different electron multiplicities.   

By integrating the rate plot, we can calculate the number of electrons appearing in this electron train from 30~$\mu$s to 1~ms, and find the typical number as a fraction of the total number of electrons produced in the primary S2. The power $b$ and amplitude $A$ have significant covariance, and the power law is not a perfect fit. Forcing the typical power $b$, the amplitude $A$ is analogous to the typical number of electrons per electron-multiplicity population in the electron train. Our goal is to reduce this background, so we compare this fraction of measured signals normalized by the primary interaction size, for different detector operating conditions.
%%
%%
%% NEXT SECTION
%%
%%
\subsection{Effect of Extraction Field}

Keeping the cathode and gate at the default biases, data was taken while varying the anode voltage from the default +5~kV down to +1~kV in 1~kV steps. A 15-minute data set was taken at each anode voltage, allowing the detector to settle for 10~minutes after changing the bias voltage and before taking data. The five corresponding extraction fields just below the liquid surface were estimated to be 5.9, 5.3, 4.7, 4.1, and 3.5~kV/cm.

If the small S2 electrons were remnants of the S2 trapped at the liquid-gas interface, we hypothesized that a better extraction efficiency from higher extraction fields would cause the electrons to be emitted faster and make the power $b$ more negative. We do not find this to be the case: the power law only significantly changes in amplitude $A$, not power $b$, despite the extraction efficiencies ranging from 50\% to 95\% ~\cite{Xu:2019dqb}.

% \begin{figure}[htb]
% \centering
% \includegraphics[scale=0.35]{ExF_frac.png}
% \caption{The number of electrons between 30 $\mu$s and 1 ms in single (green), double (blue) and 3-5 (red) electron signals as a percentage of the number of electrons produced in the primary interaction for different extraction fields (bottom x-axis) and extraction efficiencies (top x-axis). All error-bars are statistical.
% }
% \label{fig:ExF_frac}
% \end{figure}

\begin{figure}[htb]
\centering
\begin{minipage}{.5\textwidth}
  \centering
  \includegraphics[width=\linewidth]{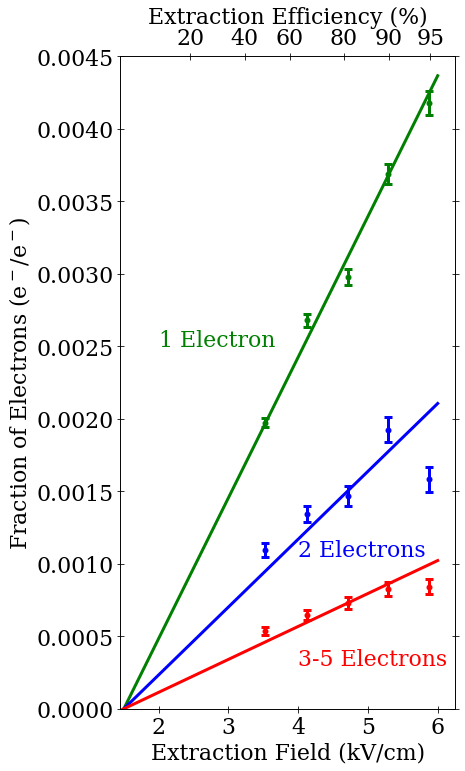}
\end{minipage}%
\begin{minipage}{.5\textwidth}
  \centering
  \includegraphics[width=\linewidth]{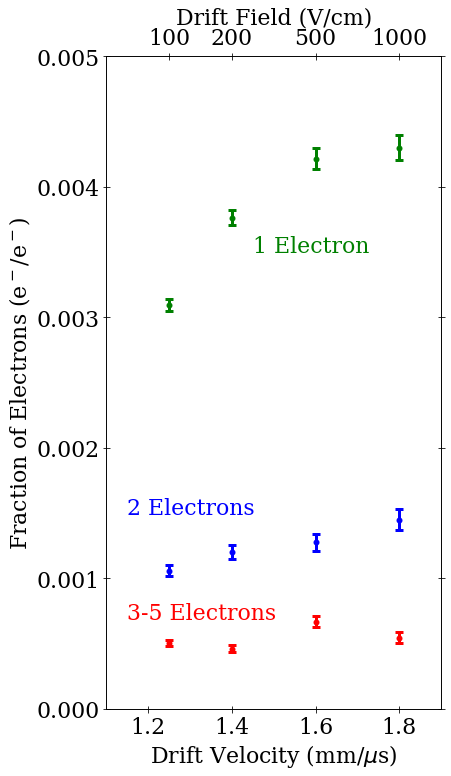}
\end{minipage}
\caption{The number of electrons between 30~$\mu$s and 1~ms in single (green), double (blue) and 3-5 (red) electron signals as a fraction of the number of electrons produced in the primary interaction, corrected for electron lifetime and drift time. The \textit{left} plot shows this fraction's dependence on different extraction fields (bottom x-axis) and electron extraction efficiencies (top x-axis). The \textit{right} plot shows this fraction's dependence on different electron drift velocities (bottom x-axis) and drift fields (top x-axis).}
\label{fig:Ex&DF}
\end{figure}

The left plot of Figure~\ref{fig:Ex&DF} shows that the number of single- and few-electron signals increases with increased extraction field and extraction efficiency. The linear trend with extraction field supports that observed by Sorensen and Kamdin in the amplitude of their slow exponential component~\cite{Sorensen:2017kpl}. The slopes of the lines of best fit are $(97 \pm 2) \cdot 10^{-5}$, $(47 \pm 3) \cdot 10^{-5}$, and $(23 \pm 1)\cdot 10^{-5}$~(kV/cm)$^{-1}$  for the single, double and 3-5 electron populations respectively. They show how the fraction of electrons in the trains relative to the number of electrons produced in the primary interaction depends on the extraction field in the liquid. Their intercept at 1.5~kV/cm is consistent with the threshold field required to extract electrons~\cite{Xu:2019dqb}. The linear effect also appears consistent with the recent findings in the PIXeY research detector, although they do not perform any fits~\cite{Bodnia:2021flk}. 

Despite different electron extraction efficiencies with different extraction fields, the rates of single and few electron signals maintain the same power $b$. However, the amplitude $A$ and therefore the total number of small signals increases directly with increased extraction field. We emphasize that the increase is linear with extraction field and not extraction efficiency.

\subsection{Effect of Drift Field}

With the anode set to the default +5~kV, the cathode bias voltage was varied between -6~kV and -5.1~kV corresponding to drift fields of 1000, 500, 200 and 100~V/cm. Again, 15-minute data sets were taken for each field, after 10-minute relaxation periods following changes to the detector conditions. 

With increased drift field, we hypothesized that the electron power law could be steepened and the fraction reduced if the dominating mechanism is electronegative impurities trapping electrons and releasing them at later times. The increased kinetic energy should make electrons less likely to become trapped or allow them to be released more quickly. For these data sets, the electron lifetime does not change with drift velocity. However, at lower fields, more electrons are lost to electronegative impurities for events near the bottom of the detector due to the longer drift times for a given distance. The power law amplitude $A$ increases with increased electron drift velocity, particularly for the single electron population, but still does not change the power $b$. The drift velocities agree with reference~\cite{Yoshino:1976zz}.

% \begin{figure}[htb]
% \centering
% \includegraphics[scale=0.35]{DrF_frac.png}
% \caption{The number of electrons between 30$\mu$s and 1ms in single (green), double (blue) and 3-5 (red) electron signals as a percentage of the number of electrons produced in the primary interaction for different electron drift velocities (bottom x-axis) and drift fields (top x-axis). All errorbars are statistical.}
% \label{fig:DrF_frac}
% \end{figure}

The right plot of Figure~\ref{fig:Ex&DF} shows that the number of single-electron signals increases with increased electron drift velocity. The number of double- and 3-5 electron signals do not exhibit a strong correlation. The measured electron lifetime in ASTERiX for all of these data sets was 3~$\mu$s, which indicates a high concentration of electronegative impurities~\cite{Aprile:2013blg,Wei:2020cwl}. Electronegative impurities are expected to only bind to a single electron each, and the fact that an effect is clearest in the single electron population indicates a potential effect from purity. 

With faster drift speeds, the measured S2 size is slightly larger, due to fewer electrons lost to the lifetime and a larger total charge yield. Therefore, the typical electron lifetime correction applied to the measured S2 size to calculate the number of electrons produced in the initial interaction decreases with increased drift field. Removing this correction, applied to the denominator in the right plot of Figure~\ref{fig:Ex&DF}, we find rather a decrease of these relative backgrounds with drift field. This is consistent with~\cite{Bodnia:2021flk}, which had a detector with an electron lifetime longer than the maximum drift time, and did not apply electron lifetime S2 corrections. This nuance indicates that the measured size rather than the produced size of the primary interaction's S2 has an effect on the behavior of the electron trains.  

\subsection{Effect of Primary Interaction Depth}

\begin{figure}[htb]
\centering
\begin{minipage}{.5\textwidth}
  \centering
  \includegraphics[width=\linewidth]{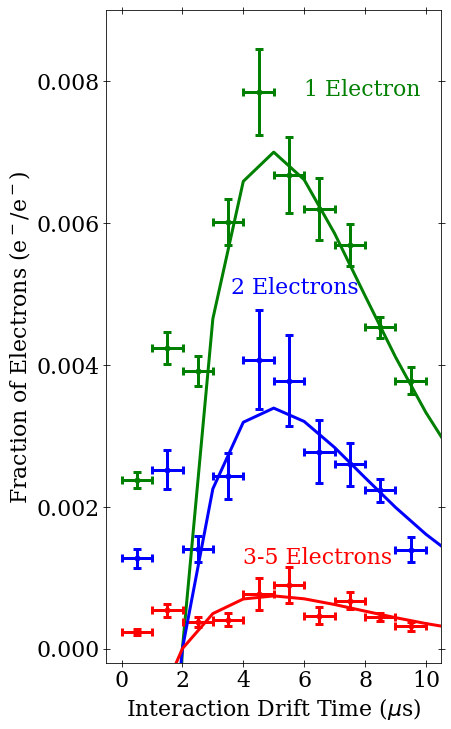}
\end{minipage}%
\begin{minipage}{.5\textwidth}
  \centering
  \includegraphics[width=\linewidth]{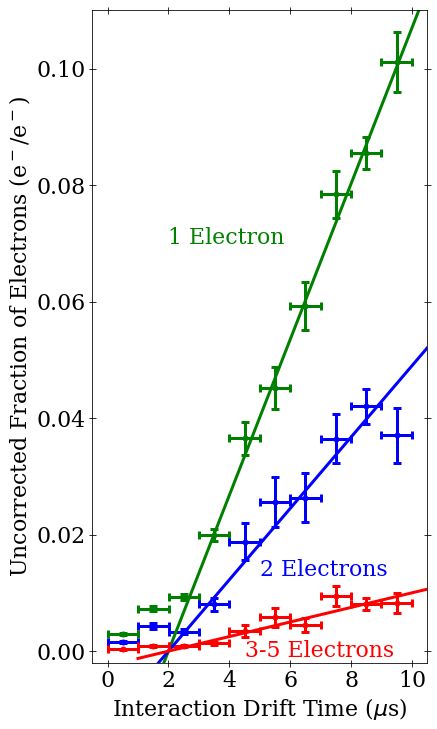}
\end{minipage}

\caption{The \textit{left} plot shows the number of electrons between 30~$\mu$s and 1~ms in single (green), double (blue) and 3-5 (red) electron signals as a fraction of the number of electrons produced in the primary interaction, corrected for electron lifetime, at different depths in the detector. The \textit{right} plot shows the same number of electron signals per population, but as a fraction of the number of measured S2 electrons, without correcting for electron lifetime. The liquid level is 2.5~mm above the gate, giving a drift time above the gate of $\sim2~\mu$s.}
\label{fig:Dt_frac}
\end{figure}

Two 15-minute data sets taken four hours apart at default detector conditions are consistent with each other, indicating the robustness of this investigation. With these higher statistics from 10,218 event windows of data, we chose $^{57}$Co events from different depths in the 1~cm detector volume below the gate. Based on the location of the source, most events were near the bottom of the target volume. The fractions of the single- and few-electron signals relative to the number of electrons produced in the primary interaction, and relative to the number of electrons measured uncorrected for lifetime, are shown in Figure~\ref{fig:Dt_frac}.

If the electron trains were dominated by delayed emission from impurities, then the fraction should always be larger at deeper positions because there are more impurities in a longer drift column to catch and release electrons. This was the observation in LUX~\cite{Akerib:2020jud}. 

If delayed emission from the liquid surface dominates, the effect of drift time on the fraction depends on the mechanism. S2s from deeper in the detector are smaller when they reach the interface because of the electron lifetime. Therefore, we expect them to be able to leave fewer electrons at the surface compared to the number of electrons produced in the original interaction and thus have smaller fractions. This appears to be the case in the left plot of Figure~\ref{fig:Dt_frac}. However, S2s from deeper in the detector are also spatially larger and less dense due to diffusion, which could affect electron emission processes of the liquid-gas interface, increasing the fraction compared to the measured S2. Diffusion area is linearly dependent on time. In the right plot of Figure~\ref{fig:Dt_frac}, the fraction is taken relative to the number of electrons left in the S2 when it reaches the surface, which is the measured S2, uncorrected for electron lifetime.

The slopes of the lines of best fit for how the fraction of trailing electrons relative to the uncorrected measured S2 depend on drift time are $(133 \pm 2) \cdot 10^{-4}$, $(61 \pm 4) \cdot 10^{-4}$, and $(12 \pm 1) \cdot 10^{-4}$~($\mu$s)$^{-1}$ for the single, double, and 3-5 electron populations respectively. The intercept at 2~$\mu$s is the location of the gate, and indicates a discontinuity consistent with the discontinuity in the electric fields and drift velocities. The left plot corrects the by the exponential electron lifetime. The lines of best fit are therefore the same but multiplied with the decreasing exponential electron lifetime factor.

This study cannot exactly disentangle whether overall smaller S2s, perhaps from less energetic interactions, cause proportionally larger trains, or whether S2s from deeper in the detector would cause larger trains. The fraction is linear rather than exponential with depth, which favors a depth dependence rather than an S2 size dependence, since the measured size of the S2 drops exponentially with drift time.

\subsection{Infrared Light-Stimulated Photodetachment}

In an effort to determine if these backgrounds were due to impurities, we irradiated the detector volume with 1~W of infrared light (1550~nm), corresponding to a photon energy of 0.8~eV. The wavelength 1550~nm is common in communication applications and therefore high powers are easily achieved. This specific data was made possible by coupling a milliwatt laser diode to an IPG Photonics EAR-1K-C-LP-SF fiber amplifier and using fiber optics and a diffuser to bring the light into the detector. A previous attempt to use \textit{in situ} infrared LEDs revealed that the components produced significantly more heat than 1550~nm light. This affected the equilibrium thermodynamics of the system, particularly since we had been using a pressurized diving bell system rather than a weir for liquid level control. We tested and confirmed that the PMTs are blind to this long wavelength, so the IR light was left on for the duration of data taking. The most common electronegative impurity is expected to be O$_2^-$, which has an electron affinity of 0.45~eV~\cite{Schiedt:1995} and a photodetachment cross-section at this photon energy of about 10$^{-19}$ cm$^2$~\cite{PhysRev.112.171}. It is likely that the relative affinity is lower and the cross-section is higher, since the ions in the detector are in an electric field. At 1~W, we have a photon density coverage of one per $\sim 5 \cdot 10^{-15}$~cm$^{2}$ in a millisecond, ignoring the high reflectivity of the PTFE walls. The relative fraction of O$_2^-$ impurities undergoing stimulated photodetachment per millisecond is conservatively estimated to be $2 \cdot 10^{-5}$. This is not as small as one might think.

We believe that it is reasonable to assume that there is a significant equilibrium concentration of O$_2^-$ built-up. The drift velocity of O$_2^-$ under the default detector conditions is 0.4~cm/s~\cite{Schmidt:2005}. By Einstein's Relation, the diffusion coefficient of O$_2^-$ is on the order of 10$^{-5}$~cm$^2$/s. The ions take seconds to reach the liquid surface, and days to diffuse from the middle to the edge of the detector. Meanwhile, tens of thousands of electrons are being lost per $^{57}$Co interaction due to the electron lifetime and tens of interactions are happening per second. Based on the number of electrons in the trains relative to the number of electrons lost to the electron lifetime, it appears that nearly all electrons caught on impurities are lost forever, and are never extracted and measured. From the electron lifetime, the O$_2$ equivalent concentration of neutral, electron-accepting impurities can be calculated~\cite{Aprile:2013blg}, which, in our case, would be about 150~ppb and approximately 10$^{17}$ impurities in the volume of ASTERiX. Therefore, despite the low photodetachment cross-section, 1~W of infrared light should significantly increase at least the position-uncorrelated single electron rate in the time region of interest from 30~$\mu$s to 1~ms. 

A typical $^{57}$Co event producing 10$^4$ electrons halfway down the detector would lose about 80\% of the electrons. But with less than $10^4$ freshly caught electrons, the infrared light could have a negligible affect on position-correlated electron train events, if they are re-emitted electrons from that primary S2. The photodetachment cross-section makes the effect of one more electron from IR light subdominant to whatever mechanism could be causing these trains. 

\begin{table}
    \centering
    \begin{tabular}{r r r }
         & Position Correlated & Position Uncorrelated \\
         &($10^{-5} e^-/e^-$ ) &($10^{-5} e^-/e^-$)\\
        
         \hline
         Single Electrons IR ON
         & 402 $\pm$ 8 
         & 127 $\pm$ 3 \\
         
         Single Electrons IR OFF
         & 416 $\pm$ 7 
         & 109 $\pm$ 3 \\
         \hline
         
         2 Electrons IR ON
         & 157 $\pm$ 7 
         & 16 $\pm$ 2 \\ 
         
         2 Electrons IR OFF
         & 141 $\pm$ 7 
         & 16 $\pm$ 2 \\
         \hline
        
         3-5 Electrons IR ON
         & 140 $\pm$ 7 
         & 13 $\pm$ 2 \\ 
         
         3-5 Electrons IR OFF
         & 129 $\pm$ 6 
         & 14 $\pm$ 2 \\
         
    \end{tabular}
    \caption{The fraction of single, double, and 3-5 electron signals from 30~$\mu$s to 1~ms after a $^{57}$Co event compared to the total number of electrons produced in the primary interaction, corrected for electron lifetime and drift time, with and without 1~W of infrared light. Position correlated events are within 9.9~mm of the primary interaction, and the uncorrelated events are outside this radius.}
    \label{tab:irtab}
\end{table}

Table~\ref{tab:irtab} lists the fraction of electrons in the signal populations relative to the number of electrons in the initial primary S2. We observe no significant effect with infrared light, except for the position-uncorrelated single electrons. This corresponds to an increase of about 20 more electrons in 1ms and could indicate an O$_2$ equivalent population of about a million in the $\sim$150~g detector. Such a number of negative ions would be produced in less than a minute in ASTERiX, so there must be a mechanism neutralizing most of these ions other than diffusion to the wall to prevent them from building up and causing significant changes to the electrodynamics and even the operation of the detector. Neutralization on the gate electrode is one of the most likely possibilities.

%% file: 4_Conclusion.tex
\section{Discussion and Conclusions}
\label{conclude}

We have investigated single- and few-electron background signals in our liquid xenon TPC that extend after an energetic interaction for times at least two orders of magnitude longer than the maximum electron drift time in the detector. These are strongly position-correlated to their primary interaction and their rates evolve as a power law with time. No detector condition that we investigated significantly altered the power law exponent $b$; changing parameters only increased or decreased the overall amplitude. 

The relative fraction of the measured electron trains compared to the number of electrons produced in the primary interaction is in total less than 1\% of those electrons produced in the interaction. The short electron lifetime of ASTERiX meant that 95\% of electrons from deep in the detector were lost while drifting. A $^{57}$Co event near the cathode would produce $\sim$ 10,000~electrons, of which only 1,000 would be detected in the primary S2, but <~50~electrons would appear in the train.

Our findings argue against the conclusion that these signals are dominated by electrons from the primary interaction that are caught and later released by impurities. Typical impurities only capture single electrons, and we observe few-electron signals at rates above what is expected by coincident single electrons. The fractions do not increase linearly with extraction efficiency, which would be expected if the electrons were caught on and released by impurities in the bulk liquid below the gate. For a constant drift field, a constant number of re-released electrons from the bulk would only be affected by the changing extraction efficiency when they reach the surface from changing the extraction field. Also, all populations increase linearly with extraction field, not just the single electrons. 

The increased fraction of the single-electrons with drift field might indicate that more electronegative impurities released their electrons at first glance. PIXeY observes an opposite trend~\cite{Bodnia:2021flk}, with which our data is only consistent when we take the fraction compared to the measured S2 rather than the produced S2 size, accounting for the electron lifetime and drift time. With an electron lifetime shorter than the maximum drift time in ASTERiX, we are able to disentangle that these electron train dependencies are more related to the measured S2 size rather than the S2 size produced in the bulk xenon drift region. This key distinction builds on previous results~\cite{Akerib:2020jud,Sorensen:2017kpl,Bodnia:2021flk}, and points to an effect at the liquid surface rather than electronegative impurities catching and releasing electrons in the drift region. The rigidity of the power $b$--despite the expectation that electronegative impurities drift faster with a higher drift field and therefore should disappear more quickly~\cite{Bodnia:2021flk}--additionally argues against an effect in the bulk.

Although more electrons are lost to electronegative impurities deeper in the detector, the fraction of electrons relative to the electrons produced in the primary does not strictly increase with increased depth of the primary interaction. Rather, the fraction of electrons in the train relative to the number of measured electrons--which is reduced according to the exponential electron lifetime with depth--increases linearly with primary interaction depth. Again, since the S2 is measured at the surface, the trend with measured S2 rather than the number of electrons produced at the interaction site points toward an effect at the surface. Attempting to induce photodetachment with infrared light does nothing, except potentially validate the theory of photodetachment by increasing position-uncorrelated single electrons. From a simple estimation of rates, it appears that an overwhelming majority of impurities that capture electrons do not release them. 

The observations that the electron backgrounds increase linearly with extraction field, that the few-electron signals cannot be coincident single electrons, that the effects depend more on measured S2 size than S2 size produced at the interaction site, and that the power law power $b$ does not change, all indicate an effect at the liquid surface. Perhaps unextracted electrons pool just below the liquid surface. Their initial cross-sectional area from their primary S2 could be determined from diffusion, which depends linearly on drift time. A surface charge density proportional to the extraction field would be expected at this dielectric surface. We might also expect a layer of electronegative impurities that have drifted to the surface, and do not have a mechanism of neutralizing, to affect the surface electrodynamics. In this case, there would be an effect of purity, but acting at the liquid surface.

The electron cloud bursting through the surface could cause mechanical ripples of the liquid that change local electric potentials. This would increase collisions in the impurity layer and/or cause points where a few electrons can be emitted from the unextracted pool or the rapidly changing surface charge density. The relaxation of the liquid ripples should not be affected by electric fields, infrared light, or where the cloud originated in the detector, and could explain the power law. A continuous sum of exponentially distributed exponentials can appear as a power law~\cite{Huntley:2006}, so damped sinusoidal ripples with exponential probabilities of electron emission with different multiplicities is a promising explanation. 

The density of these charge reservoirs would directly depend on extraction field. They could be affected by a build-up of electronegative impurities, which would be reduced with better purity. The overall charge density would directly depend on the density of the S2 when it reaches the liquid-gas interface, particularly the number of electrons and the cross-sectional area. The detector conditions determine the effect of the depth of the interaction: a smaller incident S2 would have a smaller number of electrons in the train. A larger cross-sectional area from diffusion could increase the amount of delayed, thermalized electrons at the surface as proposed by Sorensen~\cite{Sorensen:2017ymt}. Increased drift velocity could cause an increase in the electron trains, since the initial charge yield increases and the typically shorter drift times reduce the effect of the electron lifetime, both of which lead to overall larger measured S2s in primary events. Because of this, the number of electrons in the electron trains relative to the uncorrected, measured S2 size decreases with drift field. We could explain this if there was a build-up of electronegative impurities at the surface, which would have a lower equilibrium concentration if they capture electrons at a lower rate due to the increased drift field.   

In order to increase the fields, we increased the bias voltages on the stainless steel electrodes. However, due to no ``hot spots" in the full $(x,y)$-distribution of S2s, we do not believe that there was a significant emission of electrons from metal surfaces. Metal surfaces, as conductors, readily emit electrons, particularly via photoionization~\cite{Aprile:2013blg} and could be likely material origins of electrons, particularly around surface imperfections. Ultimately, we think electron emission from metals is unlikely to be the leading contribution, as it is unclear why emission processes from metals should be position-correlated with the primary interaction to such long times. Emission from the metal electrodes might still be reasonable for the position uncorrelated signals.

We are yet unable to model this theory, so the ripple hypothesis and electronegative impurity build-up at the surface remains speculation in this paper. There could also be multiple compounding processes. Our findings disagree with theories that these electron trains are from electrons caught and released by electronegative impurities in the liquid bulk, or are from metal surfaces. They rather indicate that the electron trains are an effect at the liquid-gas interface. We have thoroughly explored these backgrounds and given a clear analysis for recognizing them.

By identifying this background, we allow liquid xenon dark matter experiments to more effectively remove these backgrounds via positional and temporal cuts, based on the manifestation of the power law and position-correlation findings in a given detector. These tools also enable researchers to more carefully model these background contributions. Such a study is imperative to the success of the LBECA experiment, which aims to use a liquid xenon TPC to study low-energy interactions through few-electron ionization signals. Our characterization offers a method to improve a detector's sensitivity to such interactions, particularly from solar neutrinos and light dark matter candidates.

%% file: Acknowledgements.tex
\section*{Acknowledgements}
\label{Ack}

This work was generously supported by the Department of Energy, Office of Science under grant DE-SC0018952 and the National Science Foundation under grant PHY-1719271. Additionally, Abigail Kopec is grateful for the support of a graduate fellowship from the Indiana Space Grant Consortium. 

We recognize the valuable correspondence with Dr.~Rouven Essig, Dr.~Marivi Fernandez-Serra and Dr.~Kim Berghaus of Stonybrook regarding phenomenology and possible mechanisms. We acknowledge additional helpful discussions among experts in the LBECA Collaboration, including Dr.~Peter Sorensen, Dr.~Jingke Xu, and Dr.~Kaixuan Ni. For the electrode meshes, we thank Dr.~Petr Shagin. For the infrared light optics help and the amplifier, we thank Amy Damitz and Dr.~Dan Elliott. We appreciate the PMTs re-purposed from XENON100, and XENON's PAX framework. We also appreciate the previous manifestation of ASTERiX and initial DAQ programming made possible by Dr. Darryl Masson. We finally acknowledge the help from several temporary researchers, including Leesa Brown, Mitchell Brown, Rebecca ``Becks" Carmack, Frank DiBartolomeo, Tessa Florek, Michelle Fogel, Katrin Geigenberger, Jason Gu, Jinay Patel, Prutha Patil, Hayden Schennum, and Sowmya Seeram.

%% file: asterix.tex
\bibliographystyle{unsrt}

\bibliography{asterix.bib}